\documentclass[journal]{IEEEtran}
\normalsize
\ifCLASSINFOpdf
\else
\fi
\usepackage{microtype} \DisableLigatures[f]{encoding = *, family = * }
\usepackage{array}
\usepackage{color}
\usepackage{amsfonts}
\usepackage{amssymb}
\usepackage{amsmath}
\usepackage{graphicx}%\usepackage[dvips]{graphicx}
\usepackage{epstopdf}
\usepackage{cite}
\usepackage{balance}
\usepackage{textcomp}
\usepackage{gensymb}
\usepackage{float}
\usepackage{tabularx}
\usepackage{multirow}
\usepackage{amsthm}
\usepackage{flushend}
\usepackage[caption=false]{subfig}
\usepackage{nccmath}
\usepackage{mathtools}
\usepackage{lipsum}
\usepackage{parskip}
\usepackage{float}

\newtheorem{corollary}{\bf Corollary}
\newtheorem{proposition}{\bf Proposition}

\newcommand*{\rom}[1]{\expandafter\@slowromancap\romannumeral #1@}
\makeatletter
  \def\vhrulefill#1{\leavevmode\leaders\hrule\@height#1\hfill \kern\z@}
\makeatother

\begin{document}

\title{\huge On the Sum of Extended $\eta$-$\mu$ Variates with MRC Applications}
%\author{Author 1 and Author 2}
\author{{ Osamah S. Badarneh, \emph{Member, IEEE} and Fares S. Almehmadi}
\thanks{O. S. Badarneh is with the Electrical and Communication Engineering Department, School of Electrical Engineering and Information
Technology, German-Jordanian University, Amman 11180, Jordan (e-mail: Osamah.Badarneh@gju.edu.jo).}
\thanks{F. S. Almehmadi is with the Electrical Engineering Department, University of Tabuk, Tabuk, 71491, Saudi Arabia, (email: fal\_mehmadi@ut.edu.sa).}
}
\maketitle
\begin{abstract}
In this paper, the sum of $L$ independent but not necessarily identically distributed (i.n.i.d.) extended $\eta$-$\mu$ variates is considered. In particular, novel expressions for the probability density function and cumulative distribution function are derived in closed-forms. The derived expressions are represented in two different forms, i.e., in terms of confluent multivariate hypergeometric function and general Fox's H-function. Subsequently, closed-form expressions for the outage probability and average symbol error rate are derived. Our analytical results are validated by some numerical and Monte-Carlo simulation results.
\end{abstract}


% \fi
\IEEEpeerreviewmaketitle
\section{Introduction}
\IEEEPARstart{O}{ne} of the important diversity techniques in wireless communication systems is the maximal-ratio combining (MRC) \cite{sim}. In this technique, the transmitted signal is received by multiple receive-antennas. Hence, the output of the receiver is the sum of all received signals at each antenna (i.e., at each branch).
The sum of different fading distributions was considered in \cite{6692669,AnsariYAK17,5288931,5529757,5345735,5766060,Paris,1556824,9312154}. In particular, the sum of independent but not necessarily identically distributed (i.n.i.d.) $\eta$-$\mu$ random variables (RVs) was considered in \cite{6692669,5288931,5529757,5345735,5766060}. While the authors in \cite{AnsariYAK17,1556824} considered the sum of i.n.i.d. Nakagami-$m$ RVs. In \cite{Paris}, the authors derived analytical expressions for the  probability density function (PDF) and cumulative distribution function (CDF) of the sum of i.n.i.d. $\kappa$-$\mu$ shadowed RVs. In the above works, two important performance metrics, namely the outage probability and average symbol error rate (SER) were considered.
Recently, the extended $\eta$-$\mu$ fading models was proposed in \cite{9185088}. In contrast to the classical $\eta$-$\mu$ distribution, the extended $\eta$-$\mu$ distribution takes into consideration the power imbalance between the in-phase and quadrature components. Thus, a more realistic and flexible distribution is obtained \cite{4231253}. As a potential application, the extended distribution can be used to characterize the fading environments in device-to-device communications in which inhomogeneous with clusters of non-circularly symmetric scattered waves exist. The analysis of the extended $\eta$-$\mu$ fading models under MRC receiver has not considered yet in the literature. Motivated by this, in this paper, we address sum of i.n.i.d. extended $\eta$-$\mu$ RVs. Then, the performance of an MRC receiver is studied in terms of outage probability and average SER. Our key contributions are as follows:
\begin{itemize}
  \item The PDF of the sum of i.n.i.d. extended $\eta$-$\mu$ RVs, in terms of confluent multivariate hypergeometric function and general Fox's H-function, are derived in closed-form.
  \item The CDF of the sum of i.n.i.d. extended $\eta$-$\mu$ RVs, in terms of confluent multivariate hypergeometric function and general Fox's H-function, are derived in closed-form.
  \item The average SER for several modulation schemes, in terms of Lauricella's multivariate hypergeometric function and general Fox's H-function, are derived in closed-form.
  \item In the high signal-to-noise ratio (SNR) regime, accurate and simple approximations for the outage probability and average SER are obtained.
  \item Based on our results, the sum of i.n.i.d. classical $\eta$-$\mu$ RVs including its inclusive distributions such as the
Nakagami-$m$ and the Nakagami-$q$ (Hoyt) distributions can be deduced.
\end{itemize}
\setlength{\parskip}{0pt}
\section{The Sum Of Extended $\eta$-$\mu$ Variates}\label{sec2}
Consider an MRC receiver with $L$-branch and operating under the extended $\eta$-$\mu$ distribution. As such, the PDF of the instantaneous SNR at the $\ell$-th branch $\gamma_{\ell}$ can be obtained with the help of \cite[Eq. (14)]{9185088} as
\begin{align}
&f_{\gamma_{\ell}}(\gamma) = {1\over\Gamma (\mu_{\ell})} \left({\mu_{\ell} \xi_{\ell}\over\overline{\gamma}_{\ell}}\left({\frac {p_{\ell}}{\eta_{\ell} }}\right)^{\frac {p_{\ell}}{1+p_{\ell}}} \right)^{\mu_{\ell} } {\gamma^{\mu_{\ell} -1}} \exp \left ({-\frac {\mu_{\ell} \xi_{\ell}}{\overline{\gamma}_{\ell}}\gamma}\right) \cr&\qquad \qquad\qquad\qquad\quad {{\displaystyle { \times \, _{1}F_{1}\left ({\frac {\mu_{\ell} p_{\ell}}{1+p_{\ell}};\mu_{\ell};\frac {\mu_{\ell} \xi_{\ell} (\eta_{\ell} -p_{\ell}) }{\eta_{\ell}\,\overline{\gamma}_{\ell}}\gamma}\right) } }},&
\end{align}
%The envelope PDF is given by \cite{9185088}
%\begin{align}
%&f_{R}(r) = \frac {2 (\mu \xi)^{\mu }}{\Gamma (\mu)} \left ({\frac {p}{\eta }}\right)^{\frac {\mu p}{1+p}} \frac {r^{2\mu -1}}{\hat {r}^{2\mu }} \exp \left ({-\frac {\mu \xi r^{2}}{\hat {r}^{2}}}\right) \cr&\qquad \qquad\qquad\qquad {{\displaystyle { \times \, _{1}F_{1}\left ({\frac {\mu p}{1+p};\mu;\frac {\mu \xi (\eta -p) r^{2}}{\eta \hat {r}^{2}}}\right) } }},&
%\end{align}
in which $\overline{\gamma}_{\ell}=\mathbb{E}[\gamma]$ represents the average SNR at the $\ell$-th branch, with $\mathbb{E}[\cdot]$ being statistical expectation. Similar to the classical $\eta$-$\mu$, the extended $\eta$-$\mu$ distribution has also two formats. For both formats, $\mu={\mu_{x}+\mu_{y}\over2}$ represents the total number of multipath clusters, where $\mu_{x}$ and $\mu_{y}$ are the number of multipath clusters of the in-phase and quadrature components, respectively. In addition, $\hat {r}^{2}={\Omega_{x}+\Omega_{y}\over2}$ represents the total mean power, whereby $\Omega_{x}$ and $\Omega_{y}$ are the mean powers of the in-phase and quadrature components, respectively, Furthermore, $\xi={1+\eta\over1+p}$, where $p$ is the ratio between the number of
multipath clusters of the in-phase and quadrature components and $\eta$ describes the relationship between in-phase and quadrature
components. In Format I, the parameter $p={\mu_{x}\over\mu_{y}}$, $\eta={\Omega_{x}\over\Omega_{y}}$, $\Omega _{x}=\frac {2 \eta \hat {r}^{2} }{1+\eta}$, $\Omega _{y}=\frac {2\hat {r}^{2}}{1+\eta}$, $\mu _{x}=\frac {2 \mu p }{1+p}$, and $\mu _{y}=\frac {2 \mu }{1+p}$. While, in Format II, $p={\mu_{x}-\mu_{y}\over\mu_{x}+\mu_{y}}$, $\eta={\Omega_{x}-\Omega_{y}\over\Omega_{x}+\Omega_{y}}$, $\Omega _{x}=(1+\eta)\hat {r}^{2}$, $\Omega _{y}=(1-\eta)\hat {r}^{2}$, $\mu _{x}=(1+p)\mu $, and $\mu _{y}=(1-p)\mu$. Note that one format can be deduced for the other by simply applying the following transformation formulas: $\eta_{1}={1+\eta_{2}\over1-\eta_{2}}$ and $p_{1}={1+p_{2}\over1-p_{2}}$, where the subscripts 1 and 2 respectively refer to Format I and Format II. The analysis in this paper considers Format I.

%
%
%\begin{align}
%%&\hspace {-.5pc}
%&F_{R}(r) = \frac {(\mu \xi)^{\mu }}{\Gamma (\mu +1)}\left ({\frac {p}{\eta }}\right)^{\frac {\mu p }{1+p}}\frac {r^{2 \mu }}{\hat {r}^{2 \mu }} \cr&\qquad \qquad {{\displaystyle { \times \, \phi _{2}\!\left ({\frac {\mu p}{1+p},\frac {\mu }{1+p};\mu \!+\!1;-\frac {p r^{2} \mu \xi }{\hat {r} ^{2} \eta },-\frac {r^{2} \mu \xi }{\hat {r} ^{2}}\!}\right), } }}&
%\end{align}
%$f_{\Upsilon}={1\over2}\sqrt{\frac{\hat{r}^2}{\overline{\gamma}\gamma}}f_{R}(\sqrt{\frac{\hat{r}^2\gamma}{\overline{\gamma}}})$
% and $F_{\Upsilon}=F_{R}(\sqrt{\frac{\hat{r}^2\gamma}{\overline{\gamma}}})$.
 \begin{proposition}{The PDF of the sum of i.n.i.d. of extended $\eta$-$\mu$ variates $\Upsilon=\displaystyle\sum_{\ell=1}^{L}\gamma_{\ell}$ is given by \eqref{pdfsum1} and \eqref{pdfmrc}, where ${\mathrm{\hat{H}}}\left[\cdot\right]$ and $\Phi_{2}^{(N)}(\cdot)$ respectively denote the general Fox H-function \cite{Buschman} and the confluent hypergeometric function of $N$ variables \cite[Ch. 1, Eq. (8)]{Srivastava}.}
   \begin{align} \label{pdfsum1}
& f_\Upsilon(\gamma)\!=\! \prod_{\ell=1}^{L}\!\!\left({\mu_{\ell} \xi_{\ell}\over\overline{\gamma}_{\ell}}\left({\frac {p_{\ell}}{\eta_{\ell} }}\right)^{\frac {p_{\ell}}{1+p_{\ell}}} \right)^{\mu_{\ell} }\!\!
{\mathrm{\hat{H}}}_{2L,2L}^{0,2L}\left[{e^{\gamma}\left \vert{ \begin{array}{c} \Theta_{L}^{(1)},\Theta_{L}^{(2)}  \\[0.15cm] \Theta_{L}^{(3)},\Theta_{L}^{(4)}\end{array}}\right. }\right],&
\end{align}
where $\Theta_{L}^{(1)}{=}\left(1-A_{1},1,B_{1}\right),\ldots,\left(1-A_{L},1,B_{L}\right)$, $\Theta_{L}^{(2)}{=}\left(1-C_{1},1,D_{1}\right),\ldots,\left(1-C_{L},1,D_{L}\right)$, $\Theta_{L}^{(3)}{=}\left(-A_{1},1,B_{1}\right),\ldots,\left(-A_{L},1,B_{L}\right)$, and $\Theta_{L}^{(4)}{=}\left(-C_{1},1,D_{1}\right),\ldots,\left(-C_{L},1,D_{L}\right)$, with $A_{\ell}=\frac{\xi_{\ell} \mu_{\ell} }{\overline{\gamma}_{\ell}}$, $B_{\ell}={\frac {\mu_{\ell} }{1+p_{\ell}}}$, $C_{\ell}=\frac{p_{\ell}\xi_{\ell} \mu_{\ell} }{\eta_{\ell}\overline{\gamma}_{\ell}}, and~D_{\ell}={\frac {\mu_{\ell}p_{\ell} }{1+p_{\ell}}}$.
% are defined as follows
%\begin{align}
% \Theta_{L}^{(1)}&{=}\left(1-A_{1},1,B_{1}\right),\ldots,\left(1-A_{L},1,B_{L}\right)&\\
%%\end{align}
%%\begin{align}
% \Theta_{L}^{(2)}&{=}\left(1-C_{1},1,D_{1}\right),\ldots,\left(1-C_{L},1,D_{L}\right)\\
%%\end{align}
%%\begin{align}
% \Theta_{L}^{(3)}&{=}\left(-A_{1},1,B_{1}\right),\ldots,\left(-A_{L},1,B_{L}\right)&\\
%%\end{align}
%%\begin{align}
% \Theta_{L}^{(4)}&{=}\left(-C_{1},1,D_{1}\right),\ldots,\left(-C_{L},1,D_{L}\right).&\\
%\end{align}
\begin{figure*}
\begin{align}\label{pdfmrc}
&f_{\Upsilon}(\gamma) =\left(\prod_{\ell = 1}^{L}\left({\mu_{\ell} \xi_{\ell}\over\overline{\gamma}_{\ell}}\left({\frac {p_{\ell}}{\eta_{\ell} }}\right)^{\frac {p_{\ell}}{1+p_{\ell}}} \right)^{\mu_{\ell} }\right){\gamma^{\sum\limits_{\ell = 1}^{L}\mu_{\ell} - 1} \over \Gamma\left(\sum\limits_{\ell = 1}^{L}\mu_{\ell}\right)}\cr & \qquad\qquad\qquad\qquad\qquad\qquad\qquad\qquad\times
\Phi_{2}^{(2L)}\left(D_{1}, \ldots, D_{L},B_{1}, \ldots,B_{L}; \sum_{\ell = 1}^{L}\mu_{\ell}; -C_{1}\gamma, \ldots, -C_{L}\gamma ,-A_{1}\gamma, \ldots, -A_{L}\gamma \right)&
\end{align}
\hrulefill
\end{figure*}
\begin{figure*}
\begin{align*}\label{cdfmrc}
&F_{\Upsilon}(\gamma) = \left(\prod_{\ell = 1}^{L}\left({\mu_{\ell} \xi_{\ell}\over\overline{\gamma}_{\ell}}\left({\frac {p_{\ell}}{\eta_{\ell} }}\right)^{\frac {p_{\ell}}{1+p_{\ell}}} \right)^{\mu_{\ell} }\right){\gamma^{\sum\limits_{\ell = 1}^{L}\mu_{\ell}} \over \Gamma\left(\hbox{1}+\sum\limits_{\ell = 1}^{L}\mu_{\ell}\right)}\cr & \qquad\qquad\qquad\qquad\qquad\qquad\quad\times
\Phi_{2}^{(2L)}\left(D_{1}, \ldots, D_{L},B_{1}, \ldots,B_{L}; 1+\sum_{\ell = 1}^{L}\mu_{\ell}; -C_{1}\gamma, \ldots, -C_{L}\gamma ,-A_{1}\gamma, \ldots, -A_{L}\gamma \right){\tag{5}}&
\end{align*}
\hrulefill
\end{figure*}
\end{proposition}
\begin{IEEEproof}
See Appendix A.1.
\end{IEEEproof}
\begin{proposition}{The CDF of the sum of i.n.i.d. of extended $\eta$-$\mu$ variates $\Upsilon=\displaystyle\sum_{\ell=1}^{L}\gamma_{\ell}$ is given by  \eqref{cdfsum1} and \eqref{cdfmrc}.}
\begin{align} \label{cdfsum1}
& F_\Upsilon(\gamma) = 1+\prod_{\ell=1}^{L}\left({\mu_{\ell} \xi_{\ell}\over\overline{\gamma}_{\ell}}\left({\frac {p_{\ell}}{\eta_{\ell} }}\right)^{\frac {p_{\ell}}{1+p_{\ell}}} \right)^{\mu_{\ell} }\cr&\qquad\qquad\times
{\mathrm{\hat{H}}}_{2L+1,2L+1}^{0,2L+1}\left[{e^{\gamma}\left \vert{ \begin{array}{c} \Theta_{L}^{(1)},\Theta_{L}^{(2)} ,(1,1,1) \\[0.15cm] \Theta_{L}^{(3)},\Theta_{L}^{(4)} ,(0,1,1)\end{array}}\right. }\right].&
\end{align}
\end{proposition}
\begin{IEEEproof}
See Appendix A.2.
\end{IEEEproof}
\begin{corollary}{The PDF and the CDF of the sum of i.i.d. of extended $\eta$-$\mu$ variates $\Upsilon=\displaystyle\sum_{\ell=1}^{L}\gamma_{\ell}$ are respectively given by}
\begin{align}\label{iidpdf}
&f_{\Upsilon}(\gamma) ={{1} \over \Gamma(\mu L)}\left(\mu\xi\left({p\over\eta}\right)^{{ p\over{1}+p}}\right)^{\mu L}\left({{1} \over \bar{\gamma}}\right)^{\mu L}\gamma^{\mu L - 1}\cr&\qquad\qquad\times\Phi_{2}\left({{\mu p L \over {1}+p}},{{\mu L\over {1}+p}}; \mu L; {-p\mu\xi\over\eta\overline{\gamma}}\gamma, {-\mu\xi\over\overline{\gamma}}\gamma\right).&
\end{align}
%\end{corollary}
%\begin{corollary}{The CDF of the sum of i.i.d. of extended $\eta$-$\mu$ variates $\Upsilon=\displaystyle\sum_{\ell=1}^{L}\gamma_{\ell}$ is given by}
\begin{align}\label{iidcdf}
&F_{\Upsilon}(\gamma) = {1 \over \Gamma\left(1 + \mu L\right)}\left(\mu\xi\left({p\over\eta}\right)^{{ p\over{1}+p}}\right)^{\mu L}\left({{1} \over \bar{\gamma}}\right)^{\mu L}\gamma^{\mu L}\cr&\qquad\times\Phi_{2}\left({{\mu p L \over {1}+p}},{{\mu L\over {1}+p}}; {1} + \mu L; {-p\mu\xi\over\eta\overline{\gamma}}\gamma, {-\mu\xi\over\overline{\gamma}}\gamma\right).&
\end{align}
\end{corollary}
\begin{IEEEproof}
See Appendix A.3.
\end{IEEEproof}
It is worth mentioning that the confluent multivariate hypergeometric function, in \eqref{pdfmrc} and \eqref{cdfmrc}, involves an $L$-fold infinite summations. Besides, it has been reported in \cite{1556824} that it is inherently difficult to be evaluated when its order becomes large. However, the PDF and CDF, in \eqref{pdfsum1} and \eqref{cdfsum1}, respectively, are represented in terms of the general Fox-H function \cite{Buschman}, which involves only one single-fold integration.
%\begin{proposition}{The PDF of the sum of i.n.i.d. of extended $\eta$-$\mu$ variates $\Upsilon=\displaystyle\sum_{\ell=1}^{L}\gamma_{\ell}$ is given by}
%\end{proposition}
%\begin{proposition}{The CDF of the sum of i.n.i.d. of extended $\eta$-$\mu$ variates $\Upsilon=\displaystyle\sum_{\ell=1}^{L}\gamma_{\ell}$ is given by}
%\end{proposition}
%\begin{figure}[t]\centering
%\hspace*{0.75in}
%	\includegraphics[trim=250 20 50 50, clip,scale=.355]{Capacity_fig_a.eps}
%	\vspace*{-8mm}
%\caption{Ergodic capacity under moderate turbulence in the high-SNR regime.}\label{caphigh}
%\end{figure}
%\begin{figure}[t]\centering
%\hspace*{0.75in}
%	\includegraphics[trim=250 20 50 50, clip,scale=.355]{Capacity_fig_b.eps}
%	\vspace*{-8mm}
%\caption{Ergodic capacity under strong turbulence in the low-SNR regime.}\label{caplow}
%\end{figure}
%Note that other special cases can be deduced from our results. For example, the classical $\eta$-$\mu$ distribution can be obtained by setting $p=1$. Whereas the Hoyt (or Nakagami-$q$) is obtained by setting $p=1$, $\mu={1\over2}$, and $\eta=q^2$. While the Nakagami-$m$ case, can be obtained by setting $\eta=p$ and $\mu=m$. However, due to space limitations, these expressions are not provided here.
%Furthermore, through the imbalance Nakagami-$m$ case, the imbalance Rayleigh case can de deduced by setting $m=1$. It is worth pointing out that the PDF and CDF of the sum imbalance Nakagami-$m$ RVs has not been reported yet in the literature. However, due to space limitations, these expressions are not provided here.
\section{Performance Analysis}
\subsection{Average Symbol Error Rate}
The average SER for different modulation schemes can be evaluated based on the CDF. That is
\begin{align}\label{ser}
  {P}_{s}={\beta\delta^{\zeta}\over\Gamma(\zeta)}\int_{0}^{\infty}\gamma^{\zeta-1}\exp{(-\delta\gamma)}F_{\Upsilon}\left({\gamma}\right){\mathrm{d}}\gamma,
\end{align}
where $(\beta, \delta, \zeta)$ are modulation-dependent parameters \cite{9281340}. Thus, BFSK: $(0.5,0.5,0.5)$, BPSK: $(0.5,1,0.5)$, 4-PSK and 4-QAM: $(1,0.5,0.5)$, rectangular $M$-QAM: $(( 2(\sqrt{M}-1) )/\sqrt{M},3/(2(M-1)),0.5)$, non-rectangular $M$-QAM: $(2,3/(2(M-1)),0.5)$, $M$-PSK: $(1,\sin^{2}(\pi/M),0.5)$, and $M$-PAM: $((M-1)/M,3/(M^{2}-1),0.5)$.
%The values of $\delta$ and $\zeta$ are summarized in Table \ref{t:mod}.
%\begin{table*}[!t]
%\caption{{The Values of $\zeta$ and $\delta$ for Different Modulation Schemes }}
%\label{t:mod}
%\centering
%\begin{tabular}{|c|c|c|}
%  \hline
%  \hline
%  % after \\: \hline or \cline{col1-col2} \cline{col3-col4} ...
%  Modulation Scheme & $\zeta$ & $\delta$ \\
%   \hline\hline
%   Coherent binary frequency-shift keying (CBFSK) & ${1\over 2}$ & ${1\over 2}$ \\[0.15cm]\hline
%   Coherent binary phase-shift keying (CBPSK) & ${1\over 2}$ & $1$ \\[0.15cm]\hline
%   Non-coherent binary frequency-shift keying (NBFSK) & $1$ & ${1\over2}$ \\[0.15cm]\hline
%   Differential binary phase-shift keying (DBPSK) & $1$ & $1$ \\[0.15cm]\hline
%   $M$-ary pulse amplitude modulation ($M$-PAM)& ${1\over2}$ & $\delta= \log_{2}(M)/8(M-1)^{2}$ \\[0.15cm]\hline
%  \hline
%  \end{tabular}
%  \\[0.15cm] \hrulefill
%\end{table*}
Substituting \eqref{cdfmrc} into \eqref{ser} yields \eqref{serexp}, which can be solved with the help of \cite[Ch. 9, Eq. (43)]{Srivastava} as in \eqref{sersol}, where $F_{D}^{(N)}(\cdot)$ is the Lauricella's hypergeometric function of $N$ variables.

\begin{figure*}
\begin{align}\label{serexp}
&{P}_{s} = {\beta\delta^{\zeta}\over\Gamma(\zeta)\Gamma\left(\hbox{1}+\sum\limits_{\ell = 1}^{L}\mu_{\ell}\right)}\left(\prod_{\ell = 1}^{L}\left(\mu_{\ell}\xi_{\ell}\left({p_{\ell}\over\eta_{\ell}}\right)^{{ p_{\ell}\over{1}+p_{\ell}}}\right)^{\mu_{\ell}}\left({\hbox{1} \over \overline{\gamma}_{\ell}}\right)^{\mu_{\ell}}\right)\int_{0}^{\infty}{\gamma^{\sum\limits_{\ell = 1}^{L}\mu_{\ell}+\zeta-1}}\exp{(-\delta\gamma)}\cr &\qquad \qquad\qquad\qquad\qquad\qquad\quad\times
\Phi_{2}^{(2L)}\left(D_{1}, \ldots, D_{L},B_{1}, \ldots,B_{L}; 1+\sum_{\ell = 1}^{L}\mu_{\ell}; -C_{1}\gamma, \ldots, -C_{L}\gamma ,-A_{1}\gamma, \ldots, -A_{L}\gamma \right)&
\end{align}
\hrulefill
\end{figure*}

 \begin{figure*}
\begin{align}\label{sersol}
&P_{s} ={\beta\,\Gamma\left(\sum\limits_{\ell = 1}^{L}\mu_{\ell}+\zeta\right)\over\delta^{\sum\limits_{\ell = 1}^{L}\mu_{\ell}}\Gamma(\zeta)\Gamma\left(\hbox{1}+\sum\limits_{\ell = 1}^{L}\mu_{\ell}\right)}\left(\prod_{\ell = 1}^{L}\left(\mu_{\ell}\xi_{\ell}\left({p_{\ell}\over\eta_{\ell}}\right)^{{ p_{\ell}\over{1}+p_{\ell}}}\right)^{\mu_{\ell}}\left({\hbox{1} \over \overline{\gamma}_{\ell}}\right)^{\mu_{\ell}}\right)\cr &
\qquad\qquad\qquad\qquad\qquad\times F_{D}^{(2L)}\Bigg(\sum\limits_{\ell = 1}^{L}\mu_{\ell}+\zeta,D_{1}, \ldots, D_{L},B_{1}, \ldots,B_{L}; \hbox{1}+\sum_{\ell = 1}^{L}\mu_{\ell}; {-C_{1}\over\delta}, \ldots, {-C_{L}\over\delta} ,{-A_{1}\over\delta}, \ldots, {-A_{L}\over\delta} \Bigg)&
\end{align}
\hrulefill
\end{figure*}
Alternatively, the average SER can be obtained in terms the general Fox H-function. To this end, representing the general Fox H-function in terms of Mellin–Barnes integral, then \eqref{bera} can be rewritten as in \eqref{bera1}.
\begin{figure*}
\begin{align} \label{bera}
&{P}_{s}={\beta\delta^{\zeta}\over\Gamma(\zeta)}\int_{0}^{\infty}\gamma^{\zeta-1}\exp{(-\delta\gamma)}{\mathrm{d}}\gamma+
{\beta\delta^{\zeta}\over\Gamma(\zeta)}\prod_{\ell=1}^{L}\left(\left(\frac{\eta_{\ell}}{ p_{\ell}}\right)^{p_{\ell}}\left(\frac {\overline{\gamma}_{\ell}}{\xi_{\ell} \mu_{\ell} }\right)^{p_{\ell}+1}\right)^{-\frac {\mu_{\ell}}{1+p_{\ell}}}\cr&\qquad\qquad\qquad\qquad\qquad\qquad\qquad\qquad\qquad\qquad\times\int_{0}^{\infty}\gamma^{\zeta-1}\exp{(-\delta\gamma)}
{\mathrm{\hat{H}}}_{2L+1,2L+1}^{0,2L+1}\left[{e^{\gamma}\left \vert{ \begin{array}{c} \Theta_{L}^{(1)},\Theta_{L}^{(2)} ,(1,1,1) \\[0.15cm] \Theta_{L}^{(3)},\Theta_{L}^{(4)} ,(0,1,1)\end{array}}\right. }\right]{\mathrm{d}}\gamma&
\end{align}
\hrulefill
\end{figure*}
The first and the second integrals, with respect to $\gamma$, in \eqref{bera1} can be solved using \cite[Eq. (3.381.4)]{i:ryz}. That is
\begin{figure*}
\begin{align} \label{bera1}
&{P}_{s}={\beta\delta^{\zeta}\over\Gamma(\zeta)}\int_{0}^{\infty}\gamma^{\zeta-1}\exp{(-\delta\gamma)}{\mathrm{d}}\gamma+
{\beta\delta^{\zeta}\over\Gamma(\zeta)}\prod_{\ell=1}^{L}\left(\left(\frac{\eta_{\ell}}{ p_{\ell}}\right)^{p_{\ell}}\left(\frac {\overline{\gamma}_{\ell}}{\xi_{\ell} \mu_{\ell} }\right)^{p_{\ell}+1}\right)^{-\frac {\mu_{\ell}}{1+p_{\ell}}}\cr&\qquad\qquad\times\frac{1}{2\pi i}\oint_{C}\!\prod_{\ell=1}^{L}\left(\frac{\Gamma^{\frac {\mu_{\ell} }{1+p_{\ell}}}\left({\frac{\xi_{\ell} \mu_{\ell} }{\overline{\gamma}_{\ell}}-s}\right){\Gamma^{\frac {\mu_{\ell} p_{\ell} }{1+p_{\ell}}}\left({\frac{p_{\ell} \xi_{\ell} \mu_{\ell} }{\eta_{\ell} \overline{\gamma}_{\ell}}-s}\right)}}{\Gamma^{\frac {\mu_{\ell} }{1+p_{\ell}}}{\left(1+{\frac{\xi_{\ell} \mu_{\ell} }{\overline{\gamma}_{\ell}}-s}\right)}{\Gamma^{\frac {\mu_{\ell} p_{\ell} }{1+p_{\ell}}}\left({1}+{\frac{p_{\ell} \xi_{\ell} \mu_{\ell}}{\eta_{\ell} \overline{\gamma}_{\ell}}-s}\right)}}\right)\frac{\Gamma(-s)}{\Gamma(1-s)} \int_{0}^{\infty}\gamma^{\zeta-1}\exp{(-(\delta+s)\gamma)}
 \, {\mathrm{d}}\gamma{\mathrm{d}}s&
\end{align}
\hrulefill
\end{figure*}

\begin{align}\label{1st2nd}
 & \int_{0}^{\infty} \!\!\!\gamma^{\zeta-1}\exp{(-\delta\gamma)}{\mathrm{d}}\gamma=\delta^{-\zeta}\Gamma(\zeta)\\&
   \int_{0}^{\infty} \!\!\!\gamma^{\zeta-1}e^{-(\delta+s)\gamma}{\mathrm{d}}\gamma =(\delta+s)^{-\zeta}\Gamma(\zeta)\!\stackrel{(a)}{=} \!\frac{\delta^{-\zeta}\Gamma(\zeta)\Gamma^{\zeta}(1+\frac{s}{\delta})}{\Gamma^{\zeta}(2+\frac{s}{\delta})},\label{eq2}&
  %\int_{0}^{\infty}\gamma^{\zeta-1}\exp{(-(\delta+s)\gamma)}{\mathrm{d}}\gamma=(\delta+s)^{-\zeta}\Gamma(\zeta)\cr&\qquad\qquad\qquad\qquad\qquad\qquad \!\!\!\quad=\frac{\delta^{-\zeta}\Gamma(\zeta)\Gamma^{\zeta}(1+\frac{s}{\delta})}{\Gamma^{\zeta}(2+\frac{s}{\delta})},\label{eq2}&
\end{align}
where (a) is obtained using \cite[Eq. (8.331.1)]{i:ryz}. Substituting \eqref{1st2nd} and \eqref{eq2} into \eqref{bera1} and using the definition of the general Fox H-function, a closed-form expression for the average SER is obtained as
\begin{align} \label{bera2}
& {P}_{s} = \beta+\beta\prod_{\ell=1}^{L}\left({\mu_{\ell} \xi_{\ell}\over\overline{\gamma}_{\ell}}\left({\frac {p_{\ell}}{\eta_{\ell} }}\right)^{\frac {p_{\ell}}{1+p_{\ell}}} \right)^{\mu_{\ell} }\cr&\quad\times
{\mathrm{\hat{H}}}_{2L+2,2L+2}^{1,2L+1}\left[{\hbox{1}\left \vert{ \begin{array}{c} \Theta_{L}^{(1)},\Theta_{L}^{(2)} ,(1,1,1), (2,{1\over\delta},\zeta) \\[0.15cm] (1,{1\over\delta},\zeta),\Theta_{L}^{(3)},\Theta_{L}^{(4)} ,(0,1,1)\end{array}}\right. }\right],&
\end{align}
%Asymptotic BER, using \eqref{cdfsum1asy} into \eqref{ser} and then applying \cite[Eq. (3.381.4)]{i:ryz}, and after simple algebraic manipulations
%\begin{align} \label{bersum1asy}
%&{P}_{b}\approx {\hbox{1}\over\hbox{2}}+{\hbox{1}\over\hbox{2}}\prod_{\ell=1}^{L}\left(\left(\frac{\eta_{\ell}}{ p_{\ell}}\right)^{p_{\ell}}\left(\frac {\overline{\gamma}_{\ell}}{\xi_{\ell} \mu_{\ell} }\right)^{p_{\ell}+1}\right)^{-\frac {\mu_{\ell}}{1+p_{\ell}}}\cr&\qquad\qquad\times
%{\mathrm{\hat{H}}}_{2L+1,2L+1}^{0,2L+1}\left[{\hbox{1}\left \vert{ \begin{array}{c} \Xi_{L}^{(1)},\Xi_{L}^{(2)} ,(1,1,1) \\[0.15cm] \Xi_{L}^{(3)},\Xi_{L}^{(4)} ,(0,1,1)\end{array}}\right. }\right],&
%\end{align}

\subsection{Outage Probability}
The outage probability can be evaluated based on the derived CDFs in \eqref{cdfsum1} and \eqref{cdfmrc}. Hence, the outage probability is given by
\begin{align}\label{opexp}
  {\cal{OP}} = F_{\Upsilon}(\gamma_{o}),
\end{align}
where $\gamma_{o}$ is a predefined threshold value.
\section{Asymptotic Analysis}
In the high SNR regime, i.e., when $\overline{\gamma}_{\ell}\rightarrow\infty$, the outage probability and the average SER can be approximated as
%\begin{align}
%f_{\Upsilon}(\gamma) \!\approx\! \left(\prod_{\ell = 1}^{L}\left({\mu_{\ell}\xi_{\ell}\over\overline{\gamma}_{\ell}}\left({p_{\ell}\over\eta_{\ell}}\right)^{{ p_{\ell}\over{1}+p_{\ell}}}\right)^{\mu_{\ell}}\right){\gamma^{\sum\limits_{\ell = 1}^{L}\mu_{\ell} - 1} \over \Gamma\left(\sum\limits_{\ell = 1}^{L}\mu_{\ell}\right)}
%\end{align}
\begin{align}\label{asycdf}
\!\!\!\!F_{\Upsilon}(\gamma_{o}) \!\!\approx\!\!  \left(\prod_{\ell = 1}^{L}\left({\mu_{\ell}\xi_{\ell}\over\overline{\gamma}_{\ell}}\left({p_{\ell}\over\eta_{\ell}}\right)^{{ p_{\ell}\over{1}+p_{\ell}}}\right)^{\mu_{\ell}}\right){\gamma_{o}^{\sum\limits_{\ell = 1}^{L}\mu_{\ell}} \over \Gamma\left(\hbox{1}+\sum\limits_{\ell = 1}^{L}\mu_{\ell}\right)},
\end{align}
and
\begin{align}\label{berasy}
\!\!\!P_{s} ={\beta\,\Gamma\left(\sum\limits_{\ell = 1}^{L}\mu_{\ell}+\zeta\right)\over\delta^{\sum\limits_{\ell = 1}^{L}\mu_{\ell}}\Gamma(\zeta)\Gamma\left(\hbox{1}+\sum\limits_{\ell = 1}^{L}\mu_{\ell}\right)}\prod_{\ell = 1}^{L}\left({\mu_{\ell}\xi_{\ell}\over\overline{\gamma}_{\ell}}\left({p_{\ell}\over\eta_{\ell}}\right)^{{ p_{\ell}\over{1}+p_{\ell}}}\right)^{\mu_{\ell}}.
\end{align}
Note that \eqref{asycdf} and \eqref{berasy} are respectively obtained using \eqref{cdfmrc}, \eqref{sersol}, and based on the fact that $\Phi_{2}^{(N)}(\alpha_{1},\ldots,\alpha_{N},\omega,\underbrace{0,\ldots,0}_N)=1$ and  $F_{D}^{(N)}(\alpha_{1},\ldots,\alpha_{N},\omega,\underbrace{0,\ldots,0}_N)=1$. It is clear from \eqref{asycdf} and \eqref{berasy} that the diversity gain depends on the number of multipath parameter $\mu$ and the number of MRC branches $L$.
\section{Numerical And Simulation Results}\label{sec4}
\begin{figure}[t!] \centering
%\hspace*{-0.5in}
         \includegraphics[width=13pc,keepaspectratio]{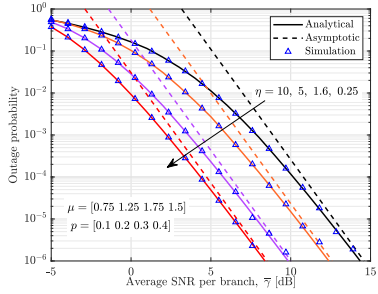}%
         \vspace*{-3mm}
       \caption{Outage probability versus average SNR per branch for quadrable-branch MRC. $\gamma_{\text{o}}=0~{\text{dB}}$, $\overline{\gamma}_{\ell}=\overline{\gamma}$ and $\eta_{\ell}=\eta$ (where $\ell=1,\ldots,4$), $\mu =  [0.75 ~1.25~ 1.75~ 1.5]$ and $p = [0.1~ 0.2~ 0.3~0.4]$.}\label{op}
\end{figure}

\begin{figure}[t!] \centering
%\hspace*{-0.5in}
         \includegraphics[width=13pc,keepaspectratio]{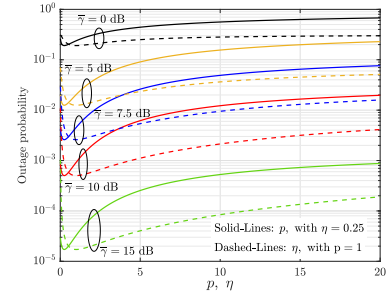}%
         \vspace*{-3mm}
       \caption{Outage probability versus $p$ and $\eta$ for i.i.d. double-branch MRC for different value of $\overline{\gamma}$ (i.e., $\overline{\gamma}=0, 5, 7.5, 10, 15$ dB) with $\gamma_{\text{o}}=0~{\text{dB}}$ and $\mu = 1.5$.}\label{opp}
\end{figure}

\begin{figure}[t!] \centering
%\hspace*{-0.5in}
         \includegraphics[width=13pc,keepaspectratio]{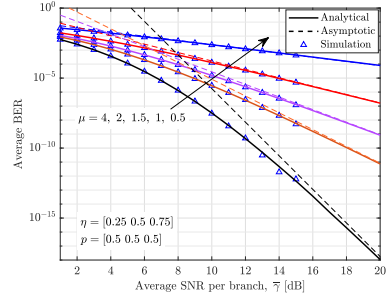}%
         \vspace*{-3mm}
       \caption{Average BER versus average SNR per branch for BPSK with triple-branch MRC. $\overline{\gamma}_{\ell}=\overline{\gamma}$ and $\mu_{\ell}=\mu$ (where $\ell=1,\ldots,3$), $\eta =  [0.25~0.5~0.75]$ and $p = [0.5~ 0.5~ 0.5]$.}\label{ber}
\end{figure}

Considering Format I, in Figs. \ref{op}-\ref{ber}, the results for outage probability and average bit error rate (BER) for BPSK under different system and channel parameters are plotted. Monte-Carlo simulation results are provided to validated the analysis. Note that efficient \textsc{Matlab} code for the functions $F_{D}$ and $\Phi_{2}$ are outlined in \cite{Butler}. While, a \textsc{Mathematica} code for the function $\hat{\text{H}}[\cdot]$ is outlined in \cite{AnsariYAK17}. of  The performance of outage probability for quadrable-branch MRC receiver under i.n.i.d. extended $\eta$-$\mu$ fading channels is depicted in Fig. \ref{op} for different values of $\eta$, while other parameters are fixed, i.e., $\gamma_{\text{o}}=0~{\text{dB}}$, $\overline{\gamma}_{\ell}=\overline{\gamma}$ (where $\ell=1,\ldots,4$), $\mu =  [0.75 ~1.25~ 1.75~ 1.5]$, and $p = [0.1~ 0.2~ 0.3~0.4]$. It can be seen form the results that the outage probability performance deteriorates as $\eta$ increases. Also, the results show that the diversity gain, which represented by the slope of asymptotic curves, is constant. This is because the diversity gain depends only on $\mu$ and(or) $L$,  as indicated by \eqref{asycdf}.
In Fig. \ref{opp}, the outage performance as a function of $\eta$ and $p$ is plotted for i.i.d. double-branch MRC receiver under different values of $\overline{\gamma}$ (i.e., $\overline{\gamma}=0, 5, 7.5, 10, 15$ dB), while $\gamma_{\text{o}}=0~{\text{dB}}$ and $\mu = 1.5$. The results demonstrate that the outage performance improves as long as $\eta\leq p$ and(or) $p\leq\eta$. The optimum performance is obtained when $\eta=p$ (which represents the Nakagami-$m$ case). When $p$ is fixed, the outage performance begins to degrades as $\eta$ becomes larger than $p$ (or as $p$ becomes larger than $\eta$ when $\eta$ is fixed.). This is due to the fact that as $p$ increases towards the value of $\eta$, the envelope PDF mode shifts rightwards, meaning a better fading
condition. Also, as $p$ becomes larger than $\eta$, the envelope PDF mode shifts back leftwards, meaning a worse fading condition. This holds for fixing $p$ and varying the value of $\eta$.

%Note that when $\eta=0.25$, when $p$ is variable, whereas $p=1$ when $\eta$ is variable. The results demonstrates that the system performance degrades as the parameters $\eta$ and $p$ increases. In particular, one can notice that the outage performance improves as $\eta$ decreases. However, this improvement {\color{blue}continues as long as} $\eta\geq p$ (i.e., $\eta\geq 1$). When $\eta$ becomes less that $p$ (i.e., $\eta<1$), the system performance deteriorates. Similarly, the system performance improves as $p$ decreases. However, this improvement {\color{blue}continues as long as} $p\geq \eta$ (i.e., $p\geq 0.25$). However, when $p$ becomes less than $\eta$ (i.e., $p<0.25$), the system performance deteriorates. Note that similar conclusion is observed over different values for $L$, $\overline{\gamma}$, $\mu$, $\eta$, and $p$. However, these results are not shown here due to space limitations.

The average BER performance for BPSK under i.n.i.d. triple-branch MRC receiver is illustrated in Fig. \ref{ber} for $\overline{\gamma}_{\ell}=\overline{\gamma}$ (where $\ell=1,\ldots,3$), $\eta =  [0.25~0.5~0.75]$, $p_{\ell} = p=0.5$ and $\mu_{\ell}=\mu$ (where $\mu = 4, 2, 1.5, 1, 0.5$). It is clear from the results that the BER performance improves as the multipath fading parameter $\mu$ increases. This is because multiple copies of the transmitted signal arrive at the receiver, which improves the performance. Also, the results show that as $\mu$ increase, the slope of the asymptotic curves (i.e., the diversity gain) becomes steeper , i.e., the diversity gain increases. In this example the diversity gain equals to $\mu L$, i.e., equals to $3\mu$.

\section{Conclusion}
The PDF and the CDF of the sum of extended $\eta$-$\mu$ variates were derived in this paper. Subsequently, the performance of an MRC receiver is analyzed in terms outage probability and average BER. The results showed that the system performance improves as number of multipath clusters, represented by the parameter $\mu$, increases. On the other hand and under i.i.d. case, the system performance degrades as the parameters $\eta$ and(or) $p$ increase(s). However, when $p$ is fixed, this degradation continues as long as $\eta> p$ (or $p>\eta$ under fixed $\eta$). When $\eta< p$ (or $p<\eta$), the system performance improves as $\eta$ increases (or $p$ increases). Finally, the results showed that the diversity gain of the system is proportional to $\mu$ and $L$.
\begin{appendices}
\section{}
 \begin{figure*}[h]
\begin{align*}\label{invpdflap}
& f_{\Upsilon}(\gamma)\! =\!{\hbox{1} \over \Gamma\left(\sum\limits_{\ell = 1}^{L}\mu_{\ell}\right)}\!\!\left(\prod_{\ell = 1}^{L}\left(\mu_{\ell}\xi_{\ell}\!\left({p_{\ell}\over\eta_{\ell}}\right)^{{ p_{\ell}\over{1}+p_{\ell}}}\right)^{\mu_{\ell}}\!\!\left({\hbox{1} \over \overline{\gamma}_{\ell}}\right)^{\mu_{\ell}}\right)\!{\cal L}^{- 1}\!\!\left[{\Gamma\left(\sum\limits_{\ell = 1}^{L}\mu_{\ell}\right) \over s^{\sum\limits_{\ell = 1}^{L} \mu_{\ell}}}\!\!\prod_{\ell = 1}^{L}\left(\hbox{1} +\frac{p_{\ell}\mu_{\ell}\xi_{\ell}}{\eta_{\ell}\overline{\gamma}_{\ell}s} \right)^{- {{\mu_{\ell} p_{\ell} \over {1}+p_{\ell}}}}  \prod_{\ell = 1}^{L}\left(\hbox{1} +\frac{\mu_{\ell}\xi_{\ell}}{\overline{\gamma}_{\ell}s} \right)^{- {{\mu_{\ell} } \over {1}+p_{\ell}}}; \gamma \right]{\tag{20}}&
\end{align*}
\hrulefill
\end{figure*}
\emph{\textbf{A.1  Proof of Proposition 1:}} The MGF of the extended $\eta$-$\mu$ distribution is given by \cite[Eq. (40)]{9185088} as
\begin{equation} \label{mfgeq1}
\mathcal{M}_{\gamma_{\ell}} (s) = \left ({1+\frac {  \overline{\gamma}_{\ell}}{\xi_{\ell} \mu_{\ell} }s}\right)^{-\frac {\mu_{\ell} }{1+p_{\ell}}}\! \left ({1+\frac { \eta_{\ell} \overline{\gamma}_{\ell}}{p_{\ell} \xi_{\ell} \mu_{\ell} }s}\right)^{-\frac {\mu_{\ell} p_{\ell} }{1+p_{\ell}}}.
\end{equation}
The PDF of $\Upsilon$ can be obtained using $f_\Upsilon(\gamma) = {\cal L}^{- 1}\left[{\cal M}_{\Upsilon}( s); \gamma\right]$, where ${\cal L}^{- 1}\left[\cdot\right]$ is the inverse Laplace transform. Thus,
 \begin{align}\label{labpdf}
 f_\Upsilon(\gamma) =\frac{1}{2\pi i}\oint_{C}\mathcal{M}_\Upsilon (s)\exp{(\gamma s)}ds,
 \end{align}
 where $\mathcal{M}_\Upsilon (s) =\prod_{\ell=1}^{L}\mathcal{M}_{\gamma_{\ell}} (s)$. Using \eqref{mfgeq1} and \eqref{labpdf}, thus \eqref{invpdflap} is obtained.

With the help of \eqref{invpdflap} and \cite[Ch. 9, Eq. (55)]{Srivastava}, a closed-form expression for the PDF of the sum of i.n.i.d. extended $\eta$-$\mu$ variates is obtained in \eqref{pdfmrc}.  However, the proof of \eqref{pdfsum1} is as follows. The MGF in \eqref{mfgeq1} can be rewritten, with the help of \cite[Eq. (8.331.1)]{i:ryz}, as
 \begin{align} \label{mfgeq}
 \setcounter{equation}{20}
&\mathcal{M}_{\gamma_{\ell}} (\gamma) =\left(\left(\frac{\eta_{\ell}}{ p_{\ell}}\right)^{p_{\ell}}\left(\frac {\overline{\gamma}_{\ell}}{\xi_{\ell} \mu_{\ell} }\right)^{p_{\ell}+1}\right)^{-\frac {\mu_{\ell}}{1+p_{\ell}}}\!\cr&\qquad\times\frac{\Gamma^{\frac {\mu_{\ell} }{1+p_{\ell}}}\left({\frac{\xi_{\ell} \mu_{\ell} }{\overline{\gamma}_{\ell}}+s}\right)}{\Gamma^{\frac {\mu_{\ell} }{1+p_{\ell}}}{\left({\frac{\xi_{\ell} \mu_{\ell} }{\overline{\gamma}_{\ell}}+s+1}\right)}}\!
\frac{\Gamma^{\frac {\mu_{\ell} p_{\ell} }{1+p_{\ell}}}\left({\frac{p_{\ell} \xi_{\ell} \mu_{\ell} }{\eta_{\ell} \overline{\gamma}_{\ell}}+s}\right)}{\Gamma^{\frac {\mu_{\ell} p_{\ell} }{1+p_{\ell}}}\left({\frac{p_{\ell} \xi_{\ell} \mu_{\ell}}{\eta_{\ell} \overline{\gamma}_{\ell}}+s+1}\right)}.&
\end{align}
%\begin{align} \label{mfgeq1}
%&\mathcal{M}_{\gamma_{\ell}} (\gamma) = \left(\frac {\overline{\gamma}_{\ell}}{\xi_{\ell} \mu_{\ell} }\right)^{-\frac {\mu_{\ell} }{1+p_{\ell}}}\!\left({\frac{\xi_{\ell} \mu_{\ell}}{\overline{\gamma}_{\ell}}+s}\right)^{-\frac {\mu_{\ell}}{1+p_{\ell}}}\! \left(\frac { \eta_{\ell} \overline{\gamma}_{\ell}}{p_{\ell} \xi_{\ell} \mu_{\ell} }\right)^{-\frac {\mu_{\ell} p_{\ell} }{1+p_{\ell}}}\cr&
%\left({\frac{p_{\ell} \xi_{\ell} \mu_{\ell} }{\eta_{\ell} \overline{\gamma}_{\ell}}+s}\right)^{-\frac {\mu_{\ell} p_{\ell} }{1+p_{\ell}}}.&
%\end{align}
% \begin{align} \label{mfgeq2}
%&\mathcal{M}_{\gamma_{\ell}} (\gamma) =\left(\left(\frac{\eta_{\ell}}{ p_{\ell}}\right)^{p_{\ell}}\left(\frac {\overline{\gamma}_{\ell}}{\xi_{\ell} \mu_{\ell} }\right)^{p_{\ell}+1}\right)^{-\frac {\mu_{\ell} }{1+p_{\ell}}}\!\left({\frac{\xi_{\ell} \mu_{\ell} }{\overline{\gamma}_{\ell}}+s}\right)^{-\frac {\mu_{\ell}}{1+p_{\ell}}}\! \cr&
%\left({\frac{p_{\ell} \xi_{\ell} \mu_{\ell} }{\eta_{\ell} \overline{\gamma}_{\ell}}+s}\right)^{-\frac {\mu_{\ell} p_{\ell} }{1+p_{\ell}}}.&
%\end{align}
Substituting \eqref{mfgeq} into \eqref{labpdf} yields
 \begin{align} \label{pdfsum}
& f_\Upsilon(\gamma) = \prod_{\ell=1}^{L}\left(\left(\frac{\eta_{\ell}}{ p_{\ell}}\right)^{p_{\ell}}\left(\frac {\overline{\gamma}_{\ell}}{\xi_{\ell} \mu_{\ell} }\right)^{p_{\ell}+1}\right)^{-\frac {\mu_{\ell}}{1+p_{\ell}}}\frac{1}{2\pi i}\!\oint_{C}\!e^{\gamma s} \cr&\!\!\times\!\prod_{\ell=1}^{L}\!\left(\frac{\Gamma^{\frac {\mu_{\ell} }{1+p_{\ell}}}\left({\frac{\xi_{\ell} \mu_{\ell} }{\overline{\gamma}_{\ell}}+s}\right){\Gamma^{\frac {\mu_{\ell} p_{\ell} }{1+p_{\ell}}}\left({\frac{p_{\ell} \xi_{\ell} \mu_{\ell} }{\eta_{\ell} \overline{\gamma}_{\ell}}+s}\right)}}{\Gamma^{\frac {\mu_{\ell} }{1+p_{\ell}}}{\left(1+{\frac{\xi_{\ell} \mu_{\ell} }{\overline{\gamma}_{\ell}}+s}\right)}{\Gamma^{\frac {\mu_{\ell} p_{\ell} }{1+p_{\ell}}}\left(1+{\frac{p_{\ell} \xi_{\ell} \mu_{\ell}}{\eta_{\ell} \overline{\gamma}_{\ell}}+s}\right)}}\right)  {\mathrm d}s, &
\end{align}
in which ${C}$ is a suitable contour in the $s$-plane. Using \eqref{pdfsum} and the Mellin-Barnes contour integral \cite[Eq. (3.1)]{Buschman}, the PDF of the sum of i.n.i.d. extended $\eta$-$\mu$ variates is obtained in a closed-form as in \eqref{pdfsum1}. This completes the proof.
%\vspace*{0.25in}
\\
\emph{\textbf{{A.2  Proof of Proposition 2:}}}
The CDF in \eqref{cdfmrc} can be obtained via $F_{\Upsilon}(\gamma)  = {\cal L}^{- 1}\left[{\cal M}_{\Upsilon}(s)/s; \gamma\right]$. With the aid of this formula, \eqref{mfgeq1}, and \cite[Ch. 9, Eq. (55)]{Srivastava}, the CDF of the sum of i.n.i.d. extended $\eta$-$\mu$ variates is obtained in a closed-form as in \eqref{cdfmrc}. However, the CDF in \eqref{cdfmrc} can be obtained by following similar steps as in \eqref{pdfsum}. Hence, the proof is completed.
%via $F_{\Upsilon}(\gamma)=1-\int_{\gamma}^{\infty}f_{\Upsilon}(\gamma){\rm{d}}\gamma$.  Thus, substituting \eqref{pdfsum} in the later formula, making change of variable $\gamma=-x$, using \cite[Eq. (3.351.2)]{i:ryz} and then applying  \cite[Eq. (8.331.1)]{i:ryz}. After that, using the definition of the general Fox's H-function, the CDF in \eqref{cdfmrc} is obtained. This completes the proof.
%\begin{figure*}
%\begin{align}\label{invcdflap}
%& F_{\Upsilon}(\gamma)  = {\hbox{1} \over \Gamma\left(\hbox{1} \!+\!\sum\limits_{\ell = 1}^{L}\mu_{\ell}\right)}\left(\prod_{\ell = 1}^{L}\left(\mu_{\ell}\xi_{\ell}\left({p_{\ell}\over\eta_{\ell}}\right)^{{ p_{\ell}\over{1}+p_{\ell}}}\right)^{\mu_{\ell}}\left({\hbox{1} \over \overline{\gamma}_{\ell}}\right)^{\mu_{\ell}}\right){\cal L}^{- 1}\left[{\Gamma\left(\hbox{1} \!+\!\sum\limits_{\ell = 1}^{L}\mu_{\ell}\right) \over s^{1+\sum\limits_{\ell = 1}^{L} \mu_{\ell}}}\prod_{\ell = 1}^{L}\left(\hbox{1} +\frac{p_{\ell}\mu_{\ell}\xi_{\ell}}{\eta_{\ell}\overline{\gamma}_{\ell}s} \right)^{- {{\mu_{\ell} p_{\ell} \over {1}+p_{\ell}}}}\right.\cr& {\hskip55pt}\left.\qquad\qquad\qquad\qquad\qquad\qquad\qquad
%\qquad\qquad\qquad\qquad\qquad\qquad\qquad\qquad\qquad\prod_{\ell = 1}^{L}\left(\hbox{1} +\frac{\mu_{\ell}\xi_{\ell}}{\overline{\gamma}_{\ell}s} \right)^{- {{\mu_{\ell} } \over {1}+p_{\ell}}}; \gamma \right]&
%\end{align}
%\hrulefill
%\end{figure*}
\\
\emph{\textbf{{A.3  Proof of Corollary 1:}}} The PDF and the CDF in \eqref{iidpdf} and \eqref{iidcdf} can be respectively obtained form \eqref{pdfmrc} and \eqref{cdfmrc} and with the help of the property of $\Phi_{2}(\cdot)$ \cite[Eq. (9)]{Paris}, which completes the proof.
\end{appendices}
%\balance
\bibliographystyle{IEEEtran}
\bibliography{FSORefe}
\onecolumn
\section{Capacity Analysis}
The average channel capacity can be obtained using
\begin{equation}\label{1}
  C = \int_{0}^{\infty}\log_{2}(1+\gamma)f_{\Gamma}(\gamma)d\gamma,\,\,\,\,\text{[bits/s/Hz]}
\end{equation}
where $f_{\Gamma}(\gamma)$ denotes the PDF of the fading model. However, using the \eqref{1} alongside the PDFs derived in the paper, there is no solution to the obtained integral. As such, to reach to a tractable solution, the logarithmic function can be approximated as [R1]
\begin{equation}\label{2}
\log_{2}(1+\gamma) \approx \sum_{k=1}^{4}\delta_{k}\exp{(-\sigma_{k}\gamma)},
\end{equation}
where $\delta=[9.331, -2.635,-4.032, -2.388]$ and $\sigma=[0.000, 0.037, 0.004,0.274]$.

Substituting \eqref{2} and the PDF in (22) into \eqref{1} yields
\begin{align} \label{pdfsum}
&C\approx \prod_{\ell=1}^{L}\left(\left(\frac{\eta_{\ell}}{ p_{\ell}}\right)^{p_{\ell}}\left(\frac {\overline{\gamma}_{\ell}}{\xi_{\ell} \mu_{\ell} }\right)^{p_{\ell}+1}\right)^{-\frac {\mu_{\ell}}{1+p_{\ell}}}\sum_{k=1}^{4}\delta_{k}\frac{1}{2\pi i}\!\oint_{C}\prod_{\ell=1}^{L}\!\left(\frac{\Gamma^{\frac {\mu_{\ell} }{1+p_{\ell}}}\left({\frac{\xi_{\ell} \mu_{\ell} }{\overline{\gamma}_{\ell}}-s}\right){\Gamma^{\frac {\mu_{\ell} p_{\ell} }{1+p_{\ell}}}\left({\frac{p_{\ell} \xi_{\ell} \mu_{\ell} }{\eta_{\ell} \overline{\gamma}_{\ell}}-s}\right)}}{\Gamma^{\frac {\mu_{\ell} }{1+p_{\ell}}}{\left(1+{\frac{\xi_{\ell} \mu_{\ell} }{\overline{\gamma}_{\ell}}-s}\right)}{\Gamma^{\frac {\mu_{\ell} p_{\ell} }{1+p_{\ell}}}\left(1+{\frac{p_{\ell} \xi_{\ell} \mu_{\ell}}{\eta_{\ell} \overline{\gamma}_{\ell}}-s}\right)}}\right)
\cr&\qquad\qquad\qquad\qquad\qquad\qquad\qquad\qquad\qquad\qquad\qquad\qquad\qquad\qquad \times\!\int_{0}^{\infty} e^{-(\sigma_{k}+s)\gamma}{\mathrm d}\gamma{\mathrm d}s.&
\end{align}
The inner integral with respect to $\gamma$ can be solved as
\begin{align}\label{1st2nd}
   \int_{0}^{\infty} e^{-(\sigma_{k}+s)\gamma}{\mathrm{d}}\gamma ={1\over(\sigma_{k}+s)}\stackrel{(a)}{=} \frac{\Gamma(\sigma_{k}+s)}{\Gamma(1+\sigma_{k}+s)},&
\end{align}
where (a) is obtained using [R2, Eq. (8.331.1)]. Substituting \eqref{1st2nd} into \eqref{pdfsum}, one can obtain
\begin{align} \label{pdfsum2}
&C\approx \prod_{\ell=1}^{L}\left(\left(\frac{\eta_{\ell}}{ p_{\ell}}\right)^{p_{\ell}}\left(\frac {\overline{\gamma}_{\ell}}{\xi_{\ell} \mu_{\ell} }\right)^{p_{\ell}+1}\right)^{-\frac {\mu_{\ell}}{1+p_{\ell}}}\sum_{k=1}^{4}\delta_{k}\cr&\times\frac{1}
{2\pi i}\!\oint_{C}\prod_{\ell=1}^{L}\!\left(\frac{\Gamma^{\frac {\mu_{\ell} }{1+p_{\ell}}}\left({\frac{\xi_{\ell} \mu_{\ell} }{\overline{\gamma}_{\ell}}-s}\right){\Gamma^{\frac {\mu_{\ell} p_{\ell} }{1+p_{\ell}}}\left({\frac{p_{\ell} \xi_{\ell} \mu_{\ell} }{\eta_{\ell} \overline{\gamma}_{\ell}}-s}\right)}}{\Gamma^{\frac {\mu_{\ell} }{1+p_{\ell}}}{\left(1+{\frac{\xi_{\ell} \mu_{\ell} }{\overline{\gamma}_{\ell}}-s}\right)}{\Gamma^{\frac {\mu_{\ell} p_{\ell} }{1+p_{\ell}}}\left(1+{\frac{p_{\ell} \xi_{\ell} \mu_{\ell}}{\eta_{\ell} \overline{\gamma}_{\ell}}-s}\right)}}\right)\frac{\Gamma(\sigma_{k}+s)}{\Gamma(1+\sigma_{k}+s)}{\mathrm d}s&
\end{align}

and use the definition of the general Fox H-function in \eqref{umgf}, the channel capacity can be obtained as
\begin{align} \label{cap}
&C\approx \prod_{\ell=1}^{L}\left(\left(\frac{\eta_{\ell}}{ p_{\ell}}\right)^{p_{\ell}}\left(\frac {\overline{\gamma}_{\ell}}{\xi_{\ell} \mu_{\ell} }\right)^{p_{\ell}+1}\right)^{-\frac {\mu_{\ell}}{1+p_{\ell}}}\sum_{k=1}^{4}\delta_{k}\,{\mathrm{\hat{H}}}_{2L+1,2L+1}^{1,2L+1}\left[{\hbox{1}\left \vert{ \begin{array}{c} \Theta_{L}^{(1)},\Theta_{L}^{(2)}, (1+\sigma_{k},1,1)  \\[0.15cm] (\sigma_{k},1,1),\Theta_{L}^{(3)},\Theta_{L}^{(4)}\end{array}}\right. }\right].&
\end{align}

Also, the average channel capacity can be obtained in terms of multivariates Lauricella's hypergeometric function as follows. Substituting \eqref{2} and the PDF in (3) (Eq. (3) in the manuscript) into \eqref{1} yields
\begin{align}\label{pdfmrc}
&C\approx\left(\prod_{\ell = 1}^{L}\left({\mu_{\ell} \xi_{\ell}\over\overline{\gamma}_{\ell}}\left({\frac {p_{\ell}}{\eta_{\ell} }}\right)^{\frac {p_{\ell}}{1+p_{\ell}}} \right)^{\mu_{\ell} }\right){1 \over \Gamma\left(\sum\limits_{\ell = 1}^{L}\mu_{\ell}\right)}\sum_{k=1}^{4}\delta_{k}\cr &\times\int_{0}^{\infty}\gamma^{\sum\limits_{\ell = 1}^{L}\mu_{\ell} - 1}\exp{(-\sigma_{k}\gamma)}
\Phi_{2}^{(2L)}\left(D_{1}, \ldots, D_{L},B_{1}, \ldots,B_{L}; \sum_{\ell = 1}^{L}\mu_{\ell}; -C_{1}\gamma, \ldots, -C_{L}\gamma ,-A_{1}\gamma, \ldots, -A_{L}\gamma \right)\mathrm{d}\gamma,\cr&
\end{align}
which can be solved as
\begin{align}\label{pdfmrc1}
&C \approx\left(\prod_{\ell = 1}^{L}\left({\mu_{\ell} \xi_{\ell}\over\overline{\gamma}_{\ell}}\left({\frac {p_{\ell}}{\eta_{\ell} }}\right)^{\frac {p_{\ell}}{1+p_{\ell}}} \right)^{\mu_{\ell} }\right)\sum_{k=1}^{4}{\delta_{k}\over\sigma_{k}^{\sum\limits_{\ell = 1}^{L}\mu_{\ell}}}\cr& \times
F_{D}^{(2L)}\left(\sum\limits_{\ell = 1}^{L}\mu_{\ell},D_{1}, \ldots, D_{L},B_{1}, \ldots,B_{L}; \sum_{\ell = 1}^{L}\mu_{\ell}; {-C_{1}\over\sigma_{k}}, \ldots, {-C_{L}\over\sigma_{k}} ,{-A_{1}\over\sigma_{k}}, \ldots, {-A_{L}\over\sigma_{k}} \right).&
\end{align}
To verify the results, Fig. \ref{cap} plots the average channel capacity for triple-branch MRC receiver. It can be noted that the analytical result is in good agreement with the Monte-Carlo simulation results. We hope that including the capacity analysis in the response letter addresses your concern.
\begin{figure}[t!] \centering
%\hspace*{-0.5in}
         \includegraphics[width=19pc,keepaspectratio]{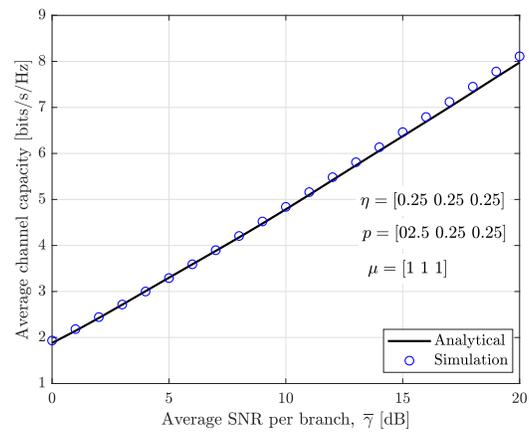}%
         \vspace*{-3mm}
       \caption{Average channel capacity versus average SNR per branch for triple-branch MRC receiver. $\overline{\gamma}_{\ell}=\overline{\gamma}$ and $\mu_{\ell}=1$, , $\eta =  0.25$, and $p = 0.25$ (where $\ell=1,2,3$).}\label{cap}
\end{figure}
\begin{enumerate}
  \item [R1.]  E. Salahat and A. Hakam, ``Novel unified expressions for error rates and ergodic channel capacity analysis over generalized fading subject to AWGGN,'' in IEEE Global Communication Conference, Austin, U.S.A, 8-12 Dec., 2014.
\end{enumerate}
\end{document}